\documentclass[useAMS,usenatbib]{mn2e}
\usepackage{graphicx}
\usepackage{rotfloat}

\newcommand{\ms}{$\,$M$_\mathrm{\odot}$}

\newcommand{\be}{\begin{equation}}
\newcommand{\ee}{\end{equation}}
\newcommand{\stars}{{\sc stars}}
\newcommand{\el}[2]{\ensuremath{^{#1}\mathrm{#2}}}

\newcommand{\an}{\ensuremath{\mathrm{(\alpha,n)}}}

\title[Mixing processes in CEMP stars]{Thermohaline mixing and gravitational settling in carbon-enhanced metal-poor stars}
\author[R.~J. Stancliffe \& E. Glebbeek]{Richard J. Stancliffe$^{1,2}$\thanks{E-mail:
Richard.Stancliffe@sci.monash.edu.au} and Evert Glebbeek$^3$\\
$^1$Institute of Astronomy, The Observatories, Madingley Road, Cambridge CB3 0HA, U.K. \\
$^2$Centre for Stellar and Planetary Astrophysics, Monash University, PO Box 28M, Clayton VIC 3800, Australia \\
$^3$Sterrekundig Instituut Utrecht, Postbus 80000, 3508 TA Utrecht, The Netherlands.
}
\begin{document}
\bibliographystyle{mn2e}

\date{Accepted 0000 December 00. Received 0000 December 00; in original form 0000 October 00}

\pagerange{\pageref{firstpage}--\pageref{lastpage}} \pubyear{0000}

\maketitle

\label{firstpage}

\begin{abstract}
We investigate the formation of carbon-enhanced metal-poor (CEMP) stars via the scenario of mass transfer from a carbon-rich asymptotic giant branch (AGB) primary to a low-mass companion in a binary system. We explore the extent to which material accreted from a companion star becomes mixed with that of the recipient, focusing on the effects of thermohaline mixing and gravitational settling. We have created a new set of asymptotic giant branch models in order to determine what the composition of material being accreted in these systems will be. We then model a range of  CEMP systems by evolving a grid of models of low-mass stars, varying the amount of material accreted by the star (to mimic systems with different separations) and also the composition of the accreted material (to mimic accretion from primaries of different mass). We find that with thermohaline mixing alone, the accreted material can become mixed with between 16 and 88\% of the pristine stellar material of the accretor, depending on the mass accreted and the composition of the material. If we include the effects of gravitational settling, we find that thermohaline mixing can be inhibited and, in the case that only a small quantity of material is accreted, can be suppressed almost completely.
\end{abstract}


\section{Introduction}
Carbon-enhanced, metal-poor (CEMP) stars are defined as stars with [C/Fe]\footnote{[A/B] = $\log (N_\mathrm{A}/N_\mathrm{B}) - \log (N_\mathrm{A}/N_\mathrm{B})_\odot$}$>$+1.0 \citep{2005ARA&A..43..531B}, with [Fe/H]$<-2$ in most cases. These objects appear with increasing frequency at low metallicity \citep{2006ApJ...652L..37L}. The study of CEMP stars is being used to probe conditions in the early universe. For example, CEMP stars have been used to infer the initial mass function in the early Galaxy \citep[e.g.][]{2005ApJ...625..833L}. Chemical abundance studies have revealed that the majority of the CEMPs are rich in $s$-process elements like barium \citep{2003IAUJD..15E..19A}, forming the so-called CEMP-s group. Recent survey work has detected a binary companion in around 68\% of these CEMP-s stars and this is consistent with them all being in binary systems \citep{2005ApJ...625..825L}.

Binary systems provide a natural explanation for these objects, which are of too low a luminosity to have been able to produce their own carbon. The primary of the system was an asymptotic giant branch (AGB) star which became carbon-rich through the action of third dredge-up\footnote{Third dredge-up occurs when the convective envelope of an AGB star deepens after a thermal pulse and material that has experienced nuclear burning is brought to the surface.} and transferred material on to the low-mass secondary (most likely via a stellar wind). The primary became a white dwarf and has long since faded from view, with the carbon-rich secondary now being the only visible component of the system. It has commonly been assumed that the accreted material remains on the surface of the secondary until the star ascends the giant branch, at which point the deepening of the convective envelope (referred to as first dredge-up because material that has experienced CN-cycling in the stellar interior is brought to the surface) mixes the material with the interior of the star. However, the transferred material should become mixed with the interior of the accreting star via the process of thermohaline mixing \citep[see e.g.][]{2007A&A...464L..57S, 2004MNRAS.355.1182C,2003NewA....8...23B}. This occurs when the mean molecular weight of the stellar gas increases toward the surface. A gas element displaced downwards and compressed will be hotter than its surroundings. It will therefore lose heat, become denser and continue to sink. This leads to mixing on thermal timescales until the molecular weight difference is eliminated \citep{1980A&A....91..175K}. \citet{2007A&A...464L..57S} showed that the inclusion of thermohaline mixing could result in the accreted material being mixed throughout 90\% of the star.

Recent work has questioned the efficiency of thermohaline mixing in carbon-enhanced metal-poor stars. Using a sample of barium-rich CEMP stars, \citet{2008ApJ...678.1351A} showed that the distribution of [C/H] values in turn-off stars (i.e. those stars that have reached the end of their main-sequence lives, are still of low-luminosity and have yet to become giants) was different from that in giants suggesting that significant mixing only happened at first dredge-up. A similar point was made by \citet{2007arXiv0709.4240D} using the data of \citet{2006ApJ...652L..37L}. These authors showed that the turn-off stars and giants were consistent with coming from the same distribution if first dredge-up resulted in the [C/H] value (and also the [N/H] value) being reduced by around 0.4 dex. They find that this result is consistent with having an accreted layer of material mixed to an average depth of about 0.2\ms\ (or alternatively having an accreted layer of 0.2\ms\ that remains unmixed).  Neither of these scenarios is consistent with the extensive mixing found by \citet{2007A&A...464L..57S}.

A possible source for reduced thermohaline mixing efficiency has been suggested by \citet{2008ApJ...677..556T}. These authors suggest that the action of gravitational settling will alter the composition gradient of the accreting star near its surface. Helium will settle from the surface, reducing the mean molecular weight at the surface but leading to an increase in the layers beneath. This produces a small region in which the mean molecular weight, $\mu$, decreases outwards toward the stellar surface -- a situation which is stable to thermohaline mixing. This stabilising composition gradient (a so-called `$\mu$-barrier') can inhibit the process of thermohaline mixing. 

This paper extends the work of \citet{2007A&A...464L..57S}, examining the effect of varying the composition of the accreted material (mimicking accretion from different masses of companion) and the amount of material that is accreted. We also investigate the effect that gravitational settling has on the extent to which material is mixed.

\section{The stellar evolution code}
Calculations in this work have been carried out using a modified version of the \stars\ stellar evolution code originally developed by \citet{1971MNRAS.151..351E} and updated by many authors \citep[e.g.][]{1995MNRAS.274..964P}. The version used here includes the nucleosynthesis routines of \citet{2005MNRAS.360..375S} and \citet{stancliffe05}, which follow the nucleosynthesis of 40 isotopes from D to \el{32}{S} and important iron group elements. The code uses the opacity routines of \citet{2004MNRAS.348..201E}, which employ interpolation in the OPAL tables \citep{1996ApJ...464..943I} and which account for the variation in opacity as the C and O content of the material varies.

At low temperatures and in the carbon-rich atmospheres of low-mass asymptotic giant branch (AGB) stars, molecular opacities become important \citep{2002A&A...387..507M}. These are not included in the \citet{2004MNRAS.348..201E} opacity routines. There are no tables of molecular opacities across a range of metallicities currently available \citep[though this is beginning to change, see][for example]{2007ApJ...667..489C},  so we adopt the procedures of \citet{2002A&A...387..507M} to account for molecular opacity. Briefly, this involves the computation of the abundances of the molecules H$_2$, H$_2$O, OH, CO, CN and C$_2$ in the equation of state and adding their contribution on to the opacity taken from the regular opacity tables. The form of the opacity for the molecules  H$_2$, H$_2$O, OH and CO are taken from \citet{1970ApJ...161..643K},  while for CN the polynomial fit of \citet{1975ApJ...200..682S} is used. The C$_2$ opacity is assumed to follow that of the CN opacity.

\section{AGB models}
We have evolved a set of models with masses of 1, 1.5, 2, 2.5, 3 and 3.5\ms\ from the pre-main-sequence without the use of convective overshooting. A mixing length of $\alpha=2.0$ has been employed throughout. This value is chosen based on calibration to a solar model. All the models were evolved using 999 mesh points. The metallicity was set to $Z=10^{-4}$ and the initial composition of the stellar material set to a solar-scaled composition \citep{1989GeCoA..53..197A}, giving the models an iron-to-hydrogen abundance of [Fe/H]$=-2.3$. Thermohaline mixing was included during the evolution and the mixing rate was enhanced by a factor of 100 above the \citet{1980A&A....91..175K} value based on the results of \citet{2007A&A...467L..15C}, who showed that this enhanced rate could reproduce the observed behaviour of certain abundance trends on the first giant branch. On the asymptotic giant branch, the AGB specific mesh spacing function of \citet{2004MNRAS.352..984S} is employed, as is the mixing scheme used by these authors. Prior to the AGB, we employ the mass-loss rate of \citet{1975MSRSL...8..369R} with $\eta=0.4$ while on the AGB we use the \citet{1993ApJ...413..641V} formula.

\begin{table}
\begin{center}
\begin{tabular}{ccccccc}
\hline
Initial & No. of & Final & Core & [C/H] & [N/H] & $\mu$ \\
mass & TPs & mass & mass &  \\
(\ms) & & (\ms) & (\ms) & &  \\
\hline
1 & 13 & 0.650 & 0.636 & 0.661 & -0.643 & 0.607 \\
1.5 & 10 & 0.892 & 0.622 & 0.613 & -1.527 & 0.621 \\
2 & 11 & 1.237 & 0.659 & 0.648 & -1.611 & 0.626 \\
2.5 & 13 & 2.422 & 0.734 & 0.501 & -1.710 & 0.619 \\
3 & 16 & 1.942 & 0.825 & 0.264 & -0.036 & 0.611 \\
3.5 & 23 & 1.871 & 0.850  & -0.030 & 0.831 & 0.628 \\
\hline
\end{tabular}
\end{center}
\caption{Details of the final state of the AGB models. The columns are: initial mass in solar masses, number of thermal pulses the model was evolved through (note that not all the envelope has been removed by the end of the run -- see the main text for more details), the final mass of the model in solar masses, the final H-exhausted core mass of the model in solar masses, the final [C/H] value, the final [N/H] value and the mean molecular weight, $\mu$, of the ejected material, assuming it is fully ionised. The values of [C/H] and [N/H] for the {\it average} composition of the ejected material will be lower.}
\label{tab:TPmodels}
\end{table}

All of the models experience third dredge-up (TDUP). The dredge-up efficiency  $\lambda$ (which is defined as the ratio of the amount of material dredged up to the growth of the core in the preceding interpulse period) is about 0.15 in the 1\ms\ model, rising to about 0.68 in the 1.5\ms\ model and is close to unity in the other models. The final outcome of the models is given in Table~\ref{tab:TPmodels}. Only the 1\ms\ model was evolved to the white dwarf cooling track. The remaining models all suffered from numerical problems in the stellar envelope during the superwind phase, with the exception of the 2.5\ms\ one which suffered an unrelated numerical problem. In each case there was a rapid increase in the stellar radius prior to the failure of the model, typically just after the peak of a thermal pulse. Similar behaviour was reported by \citet{2007PASA...24..103K}, who suggest that the problem may be related to the input physics and particularly to the opacities. Further study of this phenomenon is clearly warranted.

\begin{figure*}
\includegraphics[width=14cm, angle=270]{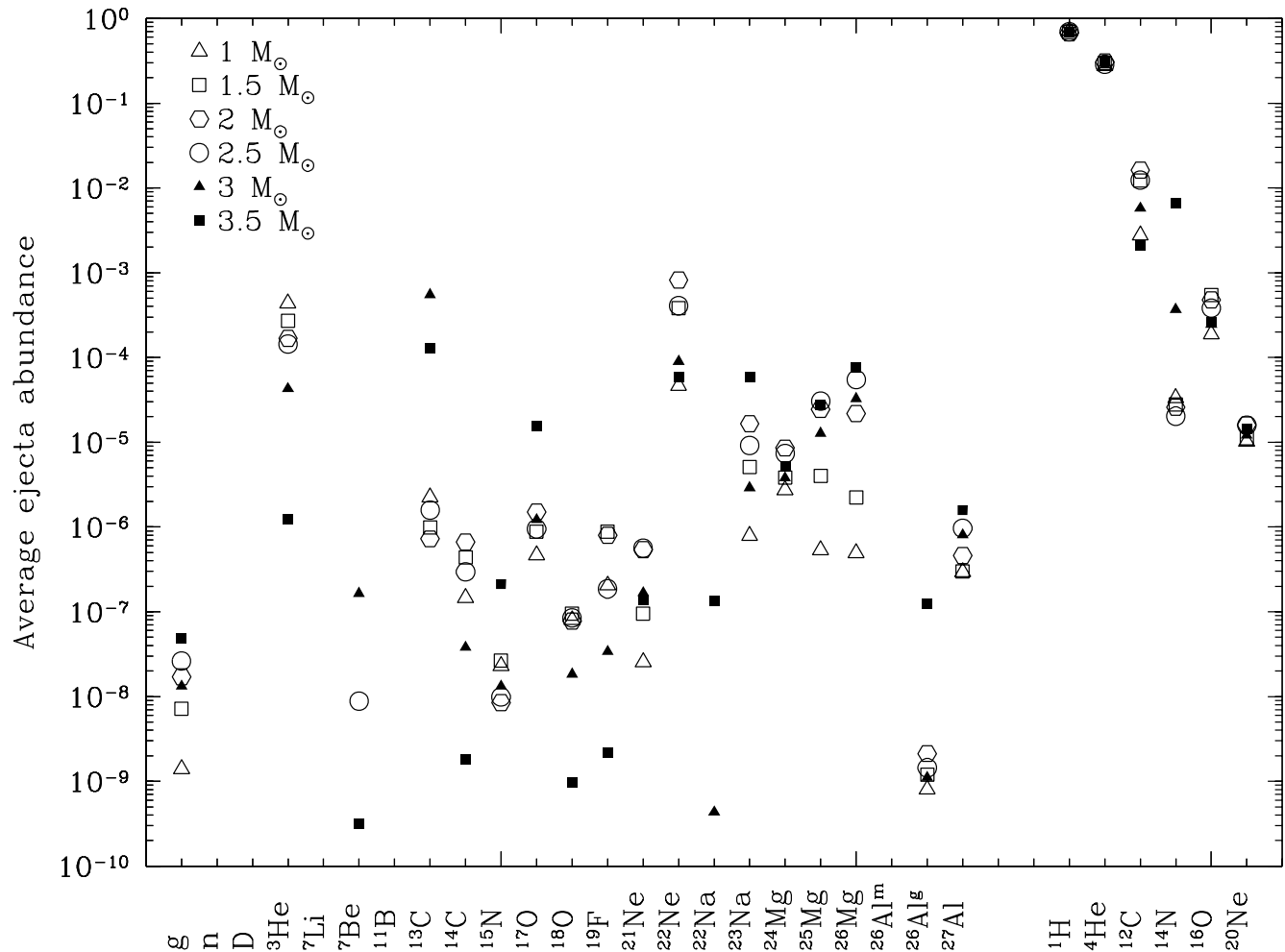}
\caption{The average abundance in the ejecta of the models for selected isotopes from the nucleosynthesis network. The elements on the right are the isotopes that are used in the structure code. The isotope `g' is a sink particle whose abundance relates to the efficiency of the s-process.}
\label{fig:avgabundance}
\end{figure*}

With the exception of the 2.5\ms\ model, all of the models have lost over half their stellar mass and they have all entered the superwind phase. We therefore estimate the contribution of the remaining envelope to the total yield by assuming that it is stripped from the star without further change to its composition. As mass loss is very rapid in the superwind phase we believe that we have missed only a few thermal pulses (and associated episodes of third dredge-up) at most. We record the final yield, the mass removed and the age of the final model for each of the masses considered. These details are needed to model the accretion on to the secondary as described below.

The average composition of material ejected from our models is displayed in Figure~\ref{fig:avgabundance}. Full details of the total mass of each isotope ejected are presented in Table~\ref{tab:agbyields} in the appendix. The trends are as one might expect for AGB stars. Below 3\ms, the models show an increase in the \el{12}{C} abundance due to the action of third dredge-up (TDUP). For the 3 and 3.5\ms\ models there is also an increase in the \el{13}{C} and \el{14}{N} abundances as hot bottom burning sets in. Hot bottom burning happens because the convective envelope is deep enough in the star to dip into the top of the H-burning shell, allowing CNO cycling to occur during the interpulse period. Similarly, there is evidence for the enhanced occurrence of the p-burning Ne-Na and Mg-Al cycles in the more massive models. The greatest abundance of \el{22}{Ne} is seen to occur in models of 1.5-2.5\ms\ masses. This isotope is produced from $\alpha$-captures on to \el{14}{N} in the intershell, where it is also destroyed by the neutron-providing reaction \el{22}{Ne}\an\el{25}{Mg}. This latter reaction becomes active at temperatures of around $3\times10^8$\,K and tends to favour destruction of \el{22}{Ne} in the hotter, more massive cores of the higher mass models. For a more detailed discussion of AGB abundance patterns and nucleosynthesis, see \citet{2007MNRAS.375.1280S} and references therein.

We now compare our models to others in the literature. Both \citet{2007PASA...24..103K} and \citet{2004ApJS..155..651H} have produced models of the same metallicity. The models of \citet{2007PASA...24..103K} -- hereafter KL07 --have similar input physics to our own, as they use similar mass-loss prescriptions and do not employ the use of convective overshooting. Herwig's models employ the \citet{1995A&A...297..727B} mass loss formula and convective overshooting is employed at both the bottom of the convective envelope and the pulse driven convection zone. 

\begin{figure}
\includegraphics[width=\columnwidth]{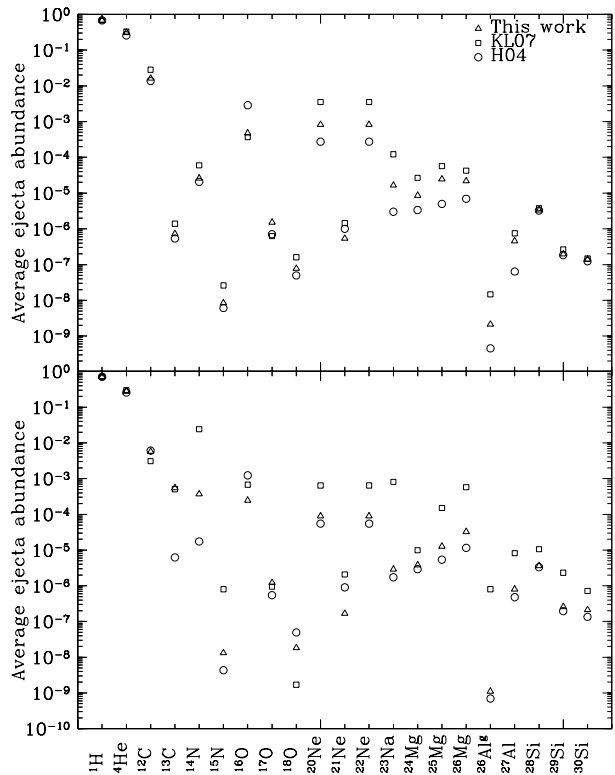}
\caption{Comparison of the average composition for isotopes in three different evolution codes. The upper panel is for a 2\ms\ model; the lower is for a 3\ms\ model. Triangles indicated the results of this work, squares are from the work of \citet{2007PASA...24..103K} -- KL07 -- and circles are the models of \citet{2004ApJS..155..651H} designated H04.}
\label{fig:ejectacomparison}
\end{figure}

KL07 have produced models of 2, 2.5, and 3\ms, along with several others that do not have the same mass as ours and hence are not considered further. \citet{2004ApJS..155..651H} -- hereafter H04 -- has produced models of 2 and 3\ms, along with higher mass models which we do not consider here. For the two initial masses common to all three works, we have plotted the average composition of the ejecta for those isotopes common to all three codes. The results are displayed in Figure~\ref{fig:ejectacomparison}. We note that there is generally good agreement between the 2\ms\ models, with few of the isotopes showing variations of over one order of magnitude. For many of the heavier isotopes, our average compositions tend to lie between those of KL07 and H04. The maximum mass dredged up after a thermal pulse (TP) is similar to that of H04, but we dredge up nearly twice the amount of material as we have nearly twice as many pulses with TDUP. Compared to KL07 we also dredge up a similar amount of material, but their model has almost twice as many TPs as ours. The number of TPs a model experiences is primarily determined by the mass loss that the model experiences. If the superwind phase begins earlier, a model will experience fewer thermal pulses. The issue of mass loss on the AGB is a serious problem for stellar models \citep[see][for a detailed discussion]{2007MNRAS.375.1280S}.

The comparison is less reassuring for the 3\ms\ models (bottom panel of Figure~\ref{fig:ejectacomparison}), with a much greater spread in abundance predictions. For example, the \el{14}{N} abundance differs by nearly 3 orders of magnitude! Again, our models tend to lie in between those of KL07 and H04. The patterns of the CNO isotopes suggests that the H04 model does not undergo significant hot bottom burning (HBB) as it has lower \el{13}{C} and \el{14}{N}, while those of this work and KL07 do. However, the KL07 model has much more \el{14}{N} (as well as \el{23}{Na} from the Ne-Na cycle) than ours due to more mass being dredged-up per pulse and there being more thermal pulses with TDUP and HBB. The KL07 model typically dredges up around $7\times10^{-3}$\ms\ per pulse (although this value reaches $1.1\times10^{-2}$ for one pulse), compared to around $5\times10^{-3}$\ms\ for our model. We also note the KL07 model has around 40 pulses with TDUP, compared to the 13 in our model.

At the low mass end, we have no direct models to compare to KL07 as we have produced 1 and 1.5\ms\ models, while they have made 1.25 and 1.75\ms\ models. However, we note similar trends. In the lower mass models, the dredge-up efficiency drops noticeably and sets of models dredge-up similar masses for each episode of TDUP. We note that once again, the KL07 models have more episodes of TDUP than ours.

\subsection{Cessation of third dredge-up}
The 1\ms\ model displays some behaviour not found in the other models, namely that TDUP operates only between pulses 3 and 8 and ceases until the final pulse, where it occurs once more.  We believe that the reason for the temporary cessation of TDUP is a consequence of the changing metallicity of the envelope. As dredge-up occurs, carbon is brought into the envelope making it more metal-rich. It is well established that stellar models of higher metallicity experience less efficient third dredge-up which tends to occur only at higher core masses and for more violent thermal pulses \citep[see e.g.][]{2002PASA...19..515K}. In the 1\ms\ model presented here, the envelope metallicity (in terms of the CNO elements) is raised up to a sufficiently high level that further thermal pulses are too weak to cause the occurrence of TDUP, until the last pulse which is significantly more violent than its predecessors.

We have also evolved a 0.9\ms\ model in the same way as described above. We find that it has an epsiode of third dredge-up on its 4$^\mathrm{th}$ thermal pulse and then TDUP ceases until the 10$^\mathrm{th}$ (and final) pulse when TDUP happens once more. We note that the pulse strengths for the last pulse is $\log L_\mathrm{He}^\mathrm{max}/\mathrm{L}_\odot = 7.84$ compared to $\log L_\mathrm{He}^\mathrm{max}/\mathrm{L}_\odot = 7.66$ for the 4$^\mathrm{th}$ pulse. Pulses 8 and 9 have pulse strengths greater than that for pulse 4, but no TDUP occurs. This seems to support the above hypothesis.

It should be noted that most authors do not find TDUP at the low envelope masses noted here \citep[see e.g.][]{2003PASA...20..389S,2007PASA...24..103K}. However, \citet{2007MNRAS.375.1280S} have found dredge-up at low envelope mass and advanced an argument why this is physically reasonable. The efficiency and occurrence of TDUP is a contentious issue and has been for some time. However, \citet{1981ASSL...88..135W} has shown that the minimum mass for dredge-up to occur is strongly dependent on the mixing length parameter, $\alpha$. He shows that for models with metallicity $Z=0.001$, the minimum stellar mass for TDUP to occur changes from 3.4\ms\ when $\alpha=0.67$ to just 1.13\ms\ when $\alpha=1.5$. We note that our $\alpha$ is larger than the value of 1.75 employed by KL07 and this could help to explain the discrepancy between computations from different codes. While this is not proof of the reason why we obtain TDUP at low envelope mass, it certainly presents an interesting avenue of further study.

\section{CEMP models}
We now turn to the modelling of the secondaries in these putative AGB binaries. We assume we are free to choose the primary mass, secondary mass and amount of material accreted to produce a secondary which has the appropriate mass (about 0.8\ms, see below). The primary mass determines the composition of the accreted material and the age at which mass is transferred to the secondary. In the case of a 1\ms\ primary, material will be transferred at an age of around $5.8\times10^9$\,years, while for a 3.5\ms\ primary matter transfer takes place at an age of around $1.9\times10^8$\, years. Variations in the amount of mass the secondary accretes would presumably be caused by variations in the separation of the system. Following \citet{2007A&A...464L..57S}, we assume that matter is accreted from the AGB wind (rather than from Roche lobe overflow). In all cases we accrete matter at a rate of $10^{-6}$\ms\,yr$^{-1}$ which is approximately the rate at which the secondary would accrete matter via the Bondi-Hoyle process \citep{1944MNRAS.104..273B} from the AGB superwind.

We evolve a grid of models of stars of 0.7, 0.75, 0.78, 0.79, 0.795, 0.798 and 0.799\ms\ each accreting enough material to make the final mass of the star equal to 0.8\ms, which is approximately the turn-off mass for the halo at the current age of the universe. We  accrete material from each of the AGB models described above on to these stars, using the average composition of the ejecta and starting the mass transfer at the final age of the AGB model. We evolve two separate sets of models: one does not include thermohaline mixing, while the other does.

At the metallicity considered here (i.e. $Z=10^{-4}$) a star of around 0.7\ms\ has only a shallow convective envelope. Any material transferred to the star does not become mixed into the stellar interior via convection. If a deep convection zone did exist (as it does in such stars at higher metallicity), accreted material would rapidly be mixed into the stellar interior by convective motions. The impact of thermohaline mixing would then be reduced as the mean molecular weight of the accreted material would decrease as it becomes diluted with the pristine stellar material in the envelope. However, at low metallicity no such convective mixing happens and thermohaline mixing may efficiently mix the accreted matter with the stellar interior.

\begin{figure}
\includegraphics[width=\columnwidth]{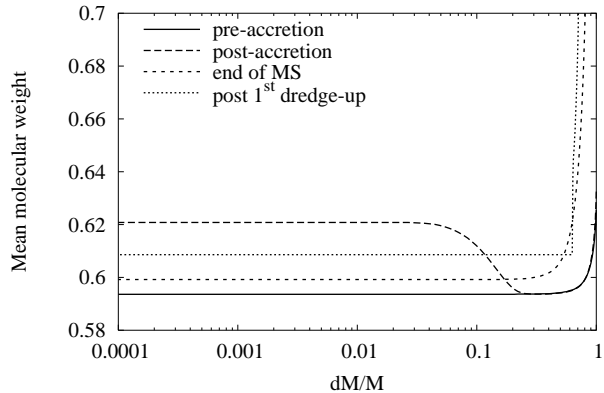}
\caption{Mean molecular weight of the stellar material as a function of the fractional mass (the mass above the considered layer) for the model accreting 0.1\ms\ of material from a 1.5\ms\ companion with thermohaline mixing included. The displayed mean molecular weight profiles are: just prior to the onset of accretion (solid line), just after accretion has ended (long-dashed line), at the end of the main sequence (short-dashed line) and just after the end of first dredge-up (dotted line). The total stellar mass has increased from 0.7\ms\ to 0.8\ms\ by the end of the accretion phase.}
\label{fig:muTH}
\end{figure}

To better understand the physical processes involved, it is instructive to consider the structural changes that occur during the evolution of one of these models. Figure~\ref{fig:muTH} shows the mean molecular weight profile at varies points in the evolution of the model which accretes 0.1\ms\ from a 1.5\ms\ companion with thermohaline mixing taken into account. Initially, the star burns hydrogen in its core, raising the mean molecular weight in the central regions while the outer regions retail their initial composition (the solid line of Figure~\ref{fig:muTH}). The star then accretes material from its companion and this material has a higher mean molecular weight than that of the original stellar material (long-dashed line). This situation is unstable to thermohaline mixing and the material mixes with the lower stellar layers, reducing the mean molecular weight of the surface layers and increasing that in the interior (short-dashed line). As this happens the star continues to burn hydrogen in the core with the mean molecular weight in this region continuing to rise. Eventually, the star evolves to the giant branch and a deep convective envelope develops. This is first dredge-up and it results in the mean molecular weight of the surface being homogenised over the whole of the convective region. The mean molecular weight of the envelope is raised because material that has experienced hydrogen burning is brought to the surface.

Table~\ref{tab:THdepth} shows the depth to which thermohaline mixing is able to mix the accreted stellar material. The most extensive mixing in the above grid mixes accreted matter with around 88\% of the accreting star. This is consistent with the work of \citet{2007A&A...464L..57S}, whose model involved accretion from a highly C-enriched 2\ms\ companion. In this work, the closest corresponding model does not mix as deeply because the companion is not as He- and C-enriched as their model was. Only 16\% of the star is mixed in the case with the shallowest mixing. The depth of mixing increases with the amount of material that is accreted. This is to be expected as the more accreted material there is, the more pristine matter must be mixed with it to reduce its mean molecular weight to something comparable to its surroundings. In addition, the lower the mass of the primary star from which material is accreted, the less efficient the mixing will be. This is because less massive stars take longer to evolve, so the accreting star has more time to burn its own material, raising the mean molecular weight of its innermost regions and preventing mixing reaching these depths.

\begin{table}
\begin{center}
\begin{tabular}{c|ccccccc}
\hline
M$_1$ & & & M$_\mathrm{acc}$ & (\ms) & \\
(\ms) & 0.1 & 0.05 & 0.02 & 0.01 & 0.005 & 0.002 & 0.001 \\
\hline
1 & 0.37 & 0.45 & 0.54 & 0.57 & 0.61 & 0.65 & 0.67\\
1.5 & 0.20 & 0.28 & 0.40 & 0.47 & 0.51 & 0.58 & 0.63\\
2 & 0.18 & 0.22 & 0.33 & 0.41 & 0.47 & 0.55 & 0.61\\
2.5 & 0.12 & 0.22 & 0.35 & 0.43 & 0.47 & 0.57 & 0.62 \\
3 & 0.15 & 0.23 & 0.35 & 0.42 & 0.50 & 0.57 & 0.62 \\
3.5 & 0.07 & 0.17 & 0.27 & 0.35 & 0.43 & 0.54 & 0.60 \\
\hline
\end{tabular}
\end{center}
\caption{Mass coordinate of the greatest extent of thermohaline mixing (in solar masses), as a function of companion mass, M$_1$, and the amount of material accreted, M$_\mathrm{acc}$. At first dredge-up, the base of the convective envelope reaches a mass co-ordinate of around 0.35\ms\ at its maximum depth.}
\label{tab:THdepth}
\end{table}

\begin{figure}
\includegraphics[width=\columnwidth]{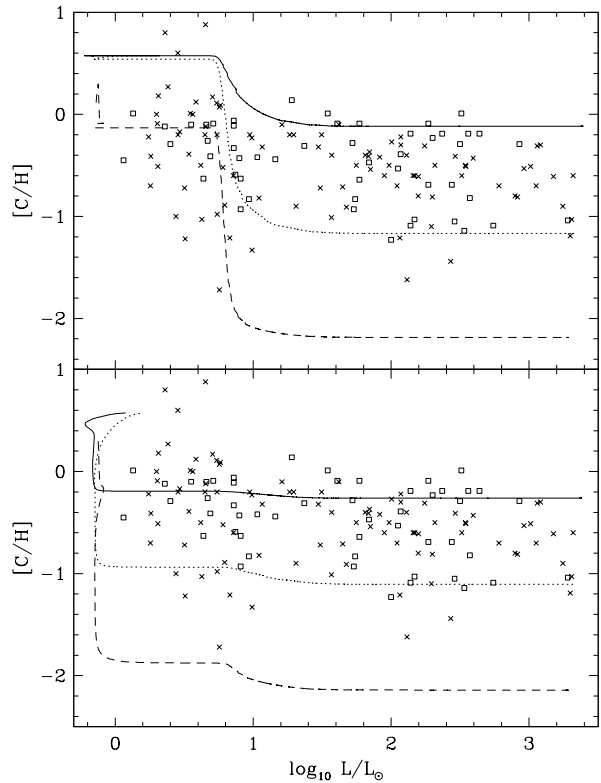}
\caption{The evolution of the surface [C/H] ratio as a function of luminosity when accreting 0.1\ms\ (solid line), 0.01\ms\ (dotted line) and 0.001\ms\ (dashed line) of material from a 1.5\ms\ companion. The mass of all the models after accretion has finished is 0.8\ms. The top panel displays models without thermohaline mixing while the lower one is for models with thermohaline mixing. The tracks begin from the point at which accretion finishes. Crosses denoted the CEMPs of \citet{2006ApJ...652L..37L} while squares denote the Ba-rich CEMPs of \citet{2007ApJ...655..492A} and \citet{2008ApJ...678.1351A}.}
\label{fig:1.5ms}
\end{figure}

\begin{figure}
\includegraphics[width=\columnwidth]{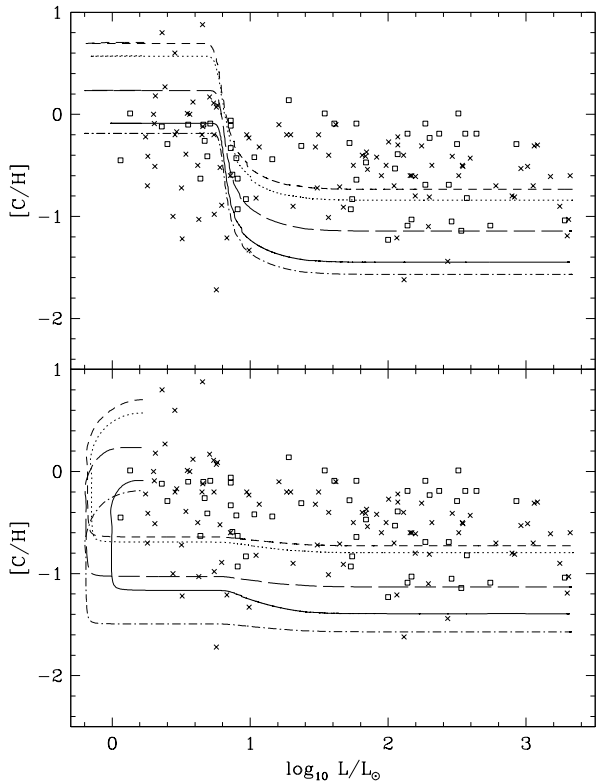}
\caption{The evolution of the surface [C/H] ratio as a function of luminosity when accreting 0.02\ms\ of material for a 1\ms\ (solid line), 1.5\ms\ (dotted line), 2\ms\ (short-dashed line), 3\ms\ (long-dashed line) and a 3.5\ms\ (dot-dashed line) companion. The mass of all the models after accretion has finished is 0.8\ms. The top panel displays models without thermohaline mixing while the lower one is for models with thermohaline mixing. The tracks begin from the point at which accretion finishes. Crosses denoted the CEMPs of \citet{2006ApJ...652L..37L} while squares denote the Ba-rich CEMPs of \citet{2007ApJ...655..492A} and \citet{2008ApJ...678.1351A}.}
\label{fig:variedcompanion}
\end{figure}

Figure~\ref{fig:1.5ms} displays the evolution of [C/H] with luminosity when accreting 0.001, 0.01 and 0.1\ms\ of material from a 1.5\ms\ companion. Note that the three models without thermohaline mixing have different [C/H] values because of the existence of a thin convective envelope, which mixes some of the material during accretion. Figure~\ref{fig:variedcompanion} displays the evolution of [C/H] with luminosity when accreting 0.02\ms\ of material from companions of different mass. These plots are representative of the full data set. 

For the models without thermohaline mixing, the only episode of mixing occurs between luminosities of around $\log L/\mathrm{L}_\odot=0.7-1.4$. This is due to the occurrence of first dredge-up as the star ascends the giant branch and a deep convective envelope develops. In some of the models there is a small amount of mixing into a surface convection zone during the accretion phase. In the models with thermohaline mixing there is a sharp drop in the [C/H] values at low luminosities while the star is on the main sequence. If a large quantity of material is accreted then it becomes mixed to a depth lower than the convective envelope reaches during first dredge-up. In this case, we do not see a decrease in [C/H] at first dredge-up. If the accreted material does not mix as deeply, when first dredge-up occurs the material becomes further diluted and the [C/H] value falls further. This can be seen in lower panel of Figure~\ref{fig:1.5ms}. The model which accretes 0.001\ms\ (the dashed line in the lower panel of this figure) only mixes down to a mass co-ordinate of 0.63\ms\ whereas the convective envelope reaches a mass co-ordinate of around 0.35\ms\ at its maximum depth. As first dredge-up occurs, we see a marked decrease in the [C/H] value as expected. 

Both sets of models have problems explaining the observed patterns of [C/H] as a function of luminosity. Models without thermohaline mixing struggle to populate those turn-off stars with low [C/H] values and unless a large quantity of material is accreted, they suffer too much dilution at first dredge-up. Models with thermohaline mixing do not produce the highest [C/H] values as the accreted matter is mixed with the pristine stellar material too quickly to be observed. In addition, they suggest that we should be able to observe turn-off stars with comparatively low [C/H] values. No such objects have so far been detected and this is not a selection effect \citep[see the discussion in ][]{2007ApJ...655..492A}. Both model sets favour accretion from 1.5-2.5\ms\ companions as accreting different masses from these stars can reproduce the spread in the [C/H] values at luminosities above $\log L/\mathrm{L}_\odot=1.4$.

\begin{figure}
\includegraphics[width=\columnwidth]{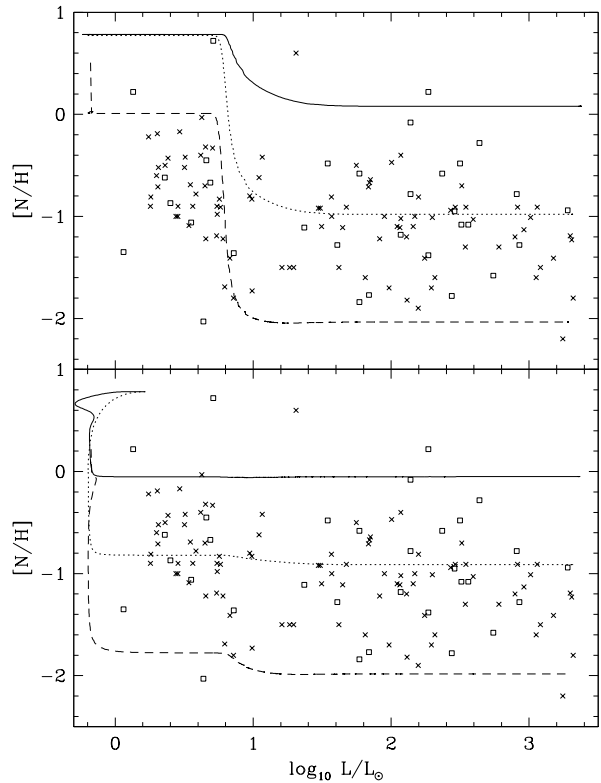}
\caption{The evolution of the surface [N/H] ratio as a function of luminosity when accreting 0.1\ms\ (solid line), 0.01\ms\ (dotted line) and 0.001\ms\ (dashed line) of material from a 3.5\ms\ companion. The mass of all the models after accretion has finished is 0.8\ms. The top panel displays models without thermohaline mixing while the lower one is for models with thermohaline mixing. The tracks begin from the point at which accretion finishes. Crosses denoted the CEMPs of \citet{2006ApJ...652L..37L} while squares denote the Ba-rich CEMPs of \citet{2007ApJ...655..492A} and \citet{2008ApJ...678.1351A}.}
\label{fig:Nfixedmass}
\end{figure}

\begin{figure}
\includegraphics[width=\columnwidth]{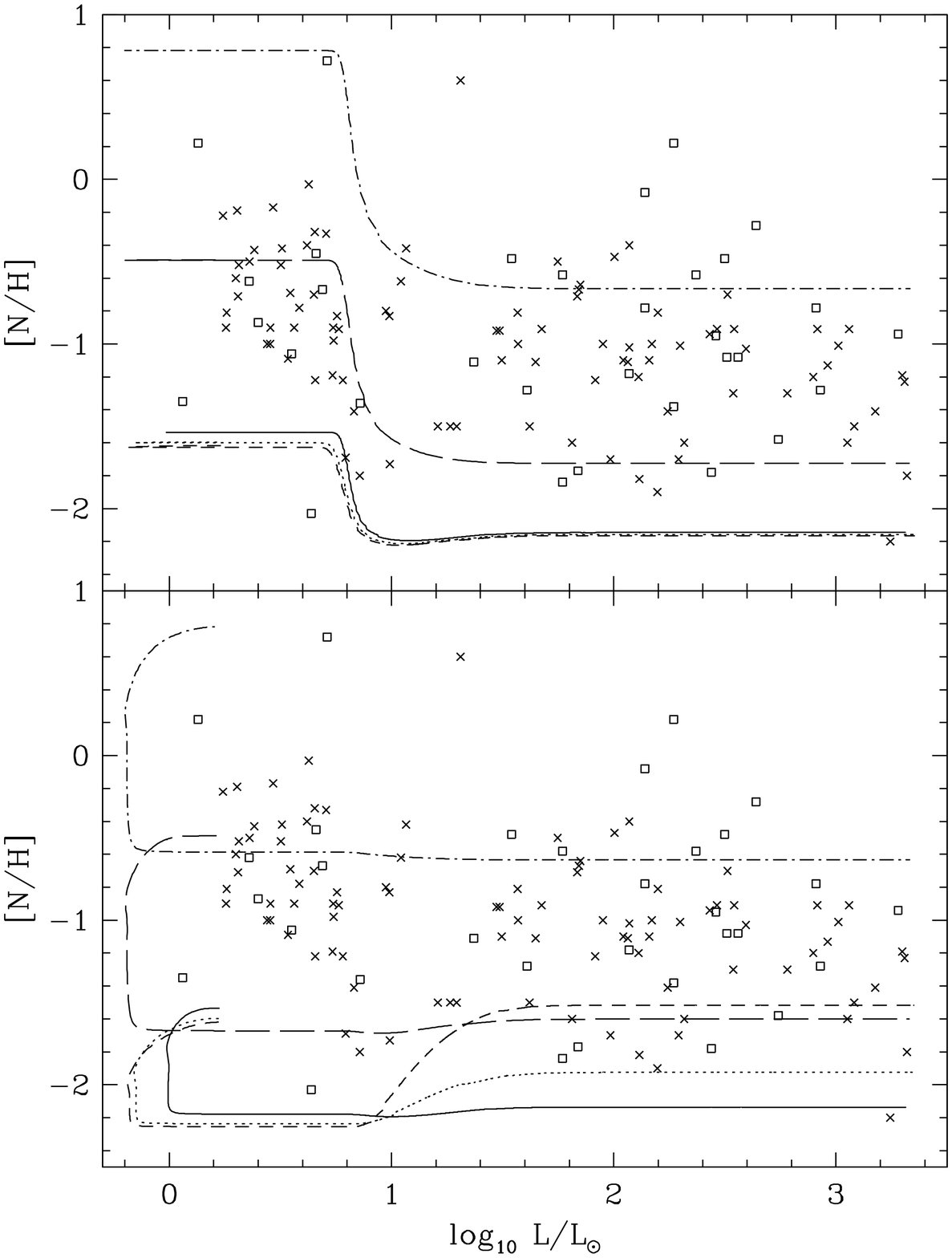}
\caption{The evolution of the surface [N/H] ratio as a function of luminosity when accreting 0.02\ms of material for a 1\ms\ (solid line), 1.5\ms\ (dotted line), 2\ms\ (short-dashed line), 3\ms\ (long-dashed line) and a 3.5\ms\ (dot-dashed line) companion. The mass of all the models after accretion has finished is 0.8\ms. The top panel displays models without thermohaline mixing while the lower one is for models with thermohaline mixing. The tracks begin from the point at which accretion finishes. Crosses denoted the CEMPs of \citet{2006ApJ...652L..37L} while squares denote the Ba-rich CEMPs of \citet{2007ApJ...655..492A} and \citet{2008ApJ...678.1351A}.}
\label{fig:Nvariedmass}
\end{figure}

The evolution of [N/H] as a function of luminosity when accreting a varying amount of mass from a companion of 3.5\ms\ is shown in Figure~\ref{fig:Nfixedmass}. Figure~\ref{fig:Nvariedmass} shows the evolution of [N/H] as a function of luminosity when accreting 0.02\ms\ of material from companions of different mass. The models without thermohaline mixing show exactly the same behaviour as was noted for [C/H], with first dredge-up causing a major dilution of the accreted material. In the case that the accreted material is nitrogen-poor (as it is for accretion from companions below 3\ms), there is a slight rise in [N/H] at the end of first-dredge-up, following the sharp drop caused by the dilution. This occurs because the convective envelope reaches down to those regions of the star where carbon originally present in the star has been processed to nitrogen via the CN cycle. This rise in [N/H] is exactly what would be expected in a `normal' metal-poor star.

The picture is very different for models including thermohaline mixing. As with the behaviour of [C/H], there is initially a sharp drop in [N/H] on the main sequence as the accreted material mixes with the stellar interior. However, as the star evolves up the giant branch (i.e. to higher luminosity), a sharp {\it increase} in [N/H] can occur during first dredge-up. This occurs because when carbon is mixed deep into the stellar interior it can be processed to nitrogen and subsequently brought to the surface during first dredge-up \citep{2007A&A...464L..57S}. This effect can be seen in the lower panel of Figure~\ref{fig:Nvariedmass}, most notably for the case of accretion from a 1.5\ms\ companion (the dashed line). The extent of this effect depends on how much carbon is present in the accreted material and to what depth it is mixed.

Nitrogen is a real problem for these models. Only the 3 and 3.5\ms\ models produce nitrogen abundances that are close to the observed values, with the 3.5\ms\ model being able to reproduce the spread in [N/H] values. Without thermohaline mixing, it is difficult to reproduce the low [N/H] values in turn-off stars, whereas the thermohaline mixing models can populate this region provided that only a small amount of material is accreted. It should be noted that such massive companions do not reproduce the [C/H] observations very well. It has been noted that it is hard to match both the carbon {\it and} nitrogen enrichments from AGB sources \citep{2007ApJ...658.1203J}.

The difficulty in matching the models to the observations is almost certainly a problem on the theoretical side. It is well known that extra-mixing mechanisms must be active during the TP-AGB (and indeed at other phases in the lives of stars). Progress at determining the physical cause of this unknown extra mixing, dubbed `cool bottom processing' (CBP) has not been forthcoming. We know that some extra-mixing processes must happen in order to get a \el{13}{C} pocket to provide neutrons for the $s$-process. However, we note that our yields are comparable to those of \citet{2004ApJS..155..651H} who included convective overshooting at the base of both the intershell convection zone and the convective envelope. It is also interesting to note that there is a degree of N-richness associated with nearly all CEMP stars. Whatever the extra-mixing mechanism is, it may have to produce just the right amount of mixing over a range of stellar masses which should seriously constrain the physical process.

One possible alternative explanation for the simultaneous enhancement of both carbon and nitrogen is the occurrence of a proton ingestion episode or flash-driven deep-mixing (FDDM). These two labels refer to the situation that occurs when protons are ingested into the intershell convection zone and burn at very high temperatures. Such episodes should produce enhancements of both C and N \citep{2000ApJ...529L..25F}. In order for the intershell convection zone to be able to penetrate into H-rich regions the hydrogen burning shell cannot present an effective entropy barrier. This only happens at very low metallicity when the efficiency of CNO burning is substantially reduced due to the absence of CNO nuclei. This does not seem to happen at [Fe/H]$\approx-2$ but may account for some of the CEMPs with lower metallicity. 

\subsection{The effect of gravitational settling}
There are certainly problems with models including thermohaline mixing from an observational perspective. They do not seem to reproduce those objects with high [C/H] ratios and they imply there should be a population of low [C/H] objects that are not observed. However from the theoretical perspective there are problems with models that do not include thermohaline mixing -- one cannot ignore physics! If thermohaline mixing does not seem to be effective, there must be some mechanism to suppress it.

\citet{2008ApJ...677..556T} have suggested that the effects of gravitational settling may inhibit the action of thermohaline mixing. Gravitational settling would lead to helium and other heavy isotopes diffusing away from the stellar surface during the main sequence. This would raise the mean molecular weight, $\mu$, just below the stellar surface, providing a positive $\mu$-gradient that would inhibit the action of thermohaline mixing \citep[see Figure 12 of ][]{2008ApJ...677..556T}. However, we note that this $\mu$-barrier is not very extensive in their model. At its peak $\mu$ is just below 0.6 which is somewhat less than the $\mu$ of the accreted material. For example, the mean molecular weight of material ejected by the 1.5\ms\ model is 0.621. This barrier is also located close to the stellar surface (in mass). One would therefore expect that the accretion of a sufficiently large amount of material will easily overcome the barrier.

To test this, we have implemented the physics of gravitational settling in the code. The rate of change of the abundance, $X_i$, of an isotope $i$ in the absence of nuclear reactions and convective mixing is given by
\be
{\partial X_i \over \partial t} = - {1\over r^2}{\partial \over \partial r} (r^2 \rho X_i v_D)
\label{eq:diffusioneq}
\ee
where $\rho$ is the density, $r$ is the radius and $v_D$ is the diffusion velocity of the species \citep{1986ApJ...307..242P}. The diffusion velocity is given by
\be
v_D = - D_{12}\left({\partial \ln c_i\over \partial r} + [m_i - \mu]{g\over kT} - \alpha_{12} {\partial \ln T\over \partial r} \right)
\ee
where $\mu$ is the mean molecular weight of the material, $g$ is the local gravity, $c_i = n(i)/n(H)$ is the concentration of the isotope $i$ relative to hydrogen and $m_i$ is the mass of the isotope $i$ \citep{1991soia.book..304M}. $D_{12}$ and $\alpha_{12}$ are the atomic and thermal diffusion coefficients and are taken from \citet{1986ApJS...61..177P}. The right-hand side of equation~\ref{eq:diffusioneq} is added to the standard equation for the composition change, which is then solved as usual.
 
\begin{figure}
\includegraphics[width=\columnwidth]{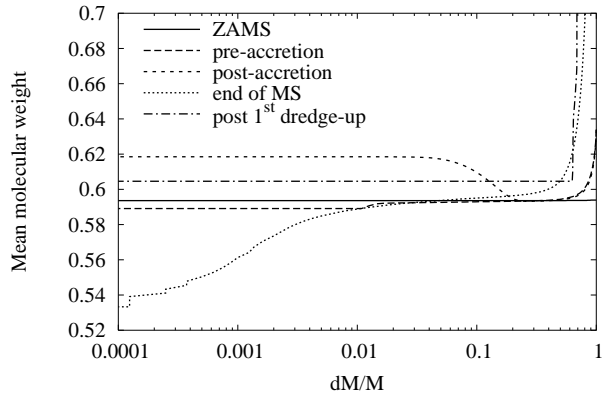}
\caption{Mean molecular weight of the stellar material as a function of the fractional mass (the mass above the considered layer) for the model accreting 0.1\ms\ of material from a 1.5\ms\ companion with thermohaline mixing and gravitational settling included. The displayed mean molecular weight profiles are: at the zero-age main sequence (ZAMS, solid line), just prior to the onset of accretion (long-dashed line), just after accretion has ended (short-dashed line), at the end of the main sequence (dotted line) and just after the end of first dredge-up (dot-dashed line). The total stellar mass has increased from 0.7\ms\ to 0.8\ms\ by the end of the accretion phase. The steps in the mean molecular weight profile near the surface of the end of main sequence model are an artefact caused by the finite precision with which the mass co-ordinate is written in the output file.}
\label{fig:muTHs}
\end{figure}

We re-ran the thermohaline mixing models with gravitational settling included in the calculation. Again, it is instructive to consider the structural changes of an individual model to understand the processes involved. Figure~\ref{fig:muTHs} shows the mean molecular weight of the interior of a 0.7\ms\ star accreting 0.1\ms\ of material from a 1.5\ms\ companion when both thermohaline mixing and gravitational settling are taken into account. At the zero-age main sequence, the composition of the material is uniform as no significant H-burning has occurred (the solid line in Figure~\ref{fig:muTHs}). Gravitational settling, which acts over long timescales, leads to helium (and other heavy elements) settling from the surface and hydrogen being enhanced there. This leads to the mean molecular weight decreasing in the surface regions (down to a fractional mass of around $\mathrm{d}M/M\approx0.01$) and a mean molecular weight gradient building up (see the long-dashed line of the figure). It is this which may potentially inhibit any thermohaline mixing. Note that the mean molecular weight remains constant down to around $\mathrm{d}M/M\approx0.01$ because of the presence of a shallow convection zone. Material of higher mean molecular weight is then accreted from the companion (short-dashed line). In this example, a large quantity of material is accreted and the mean molecular weight gradient is quickly overwhelmed. The accreted material then mixes via thermohaline mixing. Once the equilibrium configuration is reached, the effects of gravitational settling begin to assert themselves again, with helium and heavy elements becoming depleted in the surface layers by the end of the main sequence (dotted line). For this reason, the mean molecular weight in the outermost layers is reduced for a second time. Finally, the star evolves up the giant branch and first dredge-up occurs, homogenising the composition of the envelope (dot-dashed line) and producing a similar mean molecular weight profile to that of the model without gravitational settling.

Table~\ref{tab:THdepthSettling} shows the maximum extent of mixing of accreted material. We find that gravitational settling is most significant when accreting from a low mass companion. This is to be expected as settling in low-mass stars acts over long (i.e. gigayear) timescales and so it takes time to build up an effective mean molecular weight barrier. If accretion happens before the barrier is fully established, as it does in the case of accretion from all but the two lowest mass companions, then thermohaline mixing proceeds in almost the same way as it does in the absence of gravitational settling.

\begin{table}
\begin{center}
\begin{tabular}{c|ccccccc}
\hline
M$_1$ & & & M$_\mathrm{acc}$ & (\ms) & \\
(\ms) & 0.1 & 0.05 & 0.02 & 0.01 & 0.005 & 0.002 & 0.001 \\
\hline
1 & 0.43 & 0.55 & 0.69 & 0.76 & 0.78 & 0.795 & 0.797 \\
1.5 & 0.24 & 0.34 & 0.46 & 0.56 & 0.71 & 0.795 & 0.797 \\
2 & 0.16 & 0.25 & 0.39 & 0.48 & 0.56 & 0.73 & 0.794 \\
2.5 & 0.15 & 0.25 & 0.39 & 0.47 & 0.58 & 0.74 & 0.794 \\
3 &  0.17 & 0.26 & 0.38 & 0.47 & 0.58 & 0.72 & 0.791 \\
3.5 & 0.07 & 0.17 & 0.30 & 0.38 & 0.47 & 0.64 &  0.74 \\
\hline
\end{tabular}
\end{center}
\caption{Mass coordinate of the greatest extent of thermohaline mixing (in solar masses), as a function of companion mass, M$_1$, and the amount of material accreted, M$_\mathrm{acc}$ for the models including gravitational settling. At first dredge-up, the base of the convective envelope reaches a mass co-ordinate of around 0.35\ms\ at its maximum depth.}
\label{tab:THdepthSettling}
\end{table}

As the amount of material accreted is decreased, the effects of the $\mu$-barrier become more pronounced. The accreted material suffers more dilution with the H-enriched, He-depleted surface of the star which reduces its mean molecular weight, slowing the rate of thermohaline mixing. In the case that the amount of material accreted is very small, the barrier can be completely effective and mixing does not proceed to great depths. Again, this behaviour is expected. The $\mu$-barrier is formed only in the outermost parts of the star and if a large quantity of material is accreted it quickly overwhelms the barrier. When only a small amount of material is accreted the barrier is effective at inhibiting thermohaline mixing. If 0.001\ms\ of material from a 1\ms\ companion is accreted, it will only mix with less than 1\% of the star, in contrast to the 12\% seen to occur with thermohaline mixing when gravitational settling is not considered.

\begin{figure}
\includegraphics[width=\columnwidth]{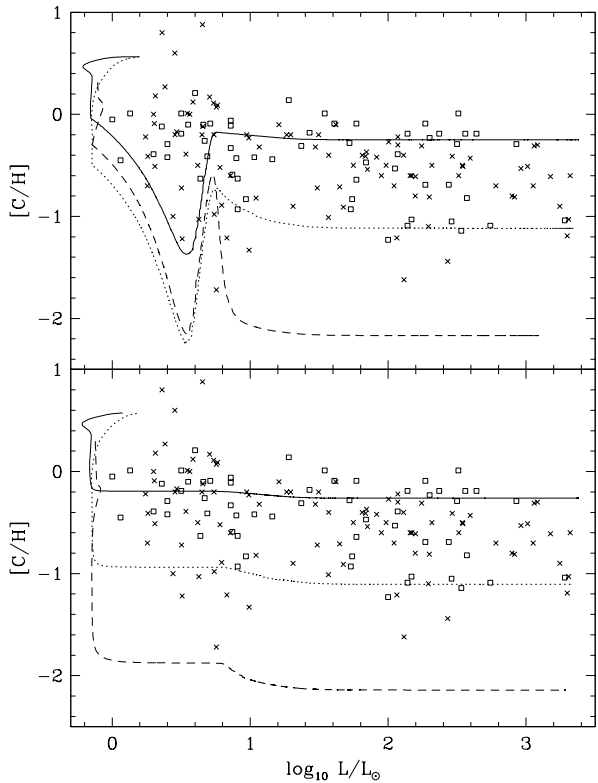}
\caption{The evolution of the surface [C/H] ratio as a function of luminosity when accreting 0.1\ms\ (solid line), 0.01\ms\ (dotted line) and 0.001\ms\ (dashed line) of material from a 1.5\ms\ companion. The mass of all the models after accretion has finished is 0.8\ms. The top panel displays models with thermohaline mixing and gravitational settling while the lower one is for models with thermohaline mixing only. The tracks begin from the point at which accretion finishes. Crosses denoted the CEMPs of \citet{2006ApJ...652L..37L} while squares denote the Ba-rich CEMPs of \citet{2007ApJ...655..492A} and \citet{2008ApJ...678.1351A}.}
\label{fig:TH+settling}
\end{figure}

This effect can also be seen in Figure~\ref{fig:TH+settling}, which displays the evolution of [C/H] as function of luminosity when material is accreted from a 1.5\ms\ companion. The decline in [C/H], followed by the sudden rise around $\log L/\mathrm{L_\odot}=0.7$ is caused by gravitational settling on the main sequence both reducing the surface carbon  abundance and enhancing that of hydrogen. The sudden rise occurs as first dredge-up sets in, with the deepening convective envelope mixing the stellar interior. When accreting 0.1\ms\ the material is quickly mixed in to the same degree in both models. If less material is accreted, we begin to notice differences in the degree of mixing. In the case of accreting 0.001\ms\ the $\mu$-barrier keeps the [C/H] ratio over 1 dex higher than in the case with thermohaline mixing alone. In fact, the $\mu$-barrier keeps the [C/H] value higher on the main sequence ($\log L/\mathrm{L_\odot}<0.7$) than it is in the case when 0.01\ms\ is accreted. This is in contrast to the situation with thermohaline mixing alone, where the model which accretes 0.001\ms\ (the dashed line in Figure~\ref{fig:TH+settling}) has a [C/H] value about 0.6 dex lower than that of the model which accretes 0.01\ms\ throughout its whole evolution.

With the inclusion of gravitational settling, we no longer obtain models with low [C/H] values at low luminosities (provided the initial [C/H] value is high enough). However, gravitational settling does present an added problem. As carbon settles from the stellar surface during the star's main-sequence lifetime, the [C/H] ratio drops. This makes it difficult for us to reproduce the more luminous turn-off objects with [C/H] values in the range -1--0. This is a serious problem for any model that includes gravitational settling. If more material is accreted, then a higher [C/H] value results throughout the main sequence. For example, if 0.2\ms\ of material were to be accreted from a 1.5\ms\ companion, the resulting object would have a [C/H] value with a minimum of just -0.5. We note that we have not included the effects of radiative levitation. However, based on the simulations of \citet{2002ApJ...568..979R} and \citet*{2002ApJ...580.1100R}, it seems that radiative levitation should be ineffective for carbon. 

There is also still the problem of the high [C/H] objects. The highest [C/H] values seem to correspond to those models which have accreted the most mass from their companions. In these cases, the $\mu$-barrier is ineffective and thermohaline mixing proceeds unhindered. The highest [C/H] value obtained in our models is about [C/H]$\ \approx-0.2$. If more material is accreted, we can reach a higher [C/H]. For example, accreting 0.2\ms\ of material from a 1.5\ms\ companion produces [C/H]$\ \approx0$. This is in line with the upper [C/H] values of the Ba-rich stars of \citet{2007ApJ...655..492A} and \citet{2008ApJ...678.1351A}. However, the sample of C-rich stars of \citet{2006ApJ...652L..37L} has three objects with [C/H]$\ >0.6$. It is difficult to explain these objects with models involving thermohaline mixing as our AGB models only just produce [C/H] values of around this. Any amount of mixing will easily reduce this value. However, we note that these three high [C/H] CEMPs are all somewhat more metal-rich (with [Fe/H] greater than around -1.8) than the models of this work so these models may not be directly comparable.

\section{Discussion}
As suggested by \citet{2008ApJ...677..556T} a means of suppressing the action of thermohaline mixing is necessary to prevent the destruction of lithium in some CEMPs. Lithium is a fragile element that is easily destroyed at temperatures of around $2\times10^6$\,K which are found at around 0.01\ms\ below the stellar surface of the low-mass secondary while on the main sequence. As we have seen, gravitational settling will only allow the accreted matter to remain at the surface if the mass of accreted material is small. In the case of CS22964-161, the CEMP binary of \citet{2008ApJ...677..556T}, the measured [C/H] value of -1.2 fits well with accretion from a 1-1.5\ms\ companion when about  0.001\ms\ is transferred to the recipient star. We would expect lithium to survive in this model because mixing does not reach down to regions where the temperature is high enough for lithium to burn. However, we have not tried to model this. We leave the treatment of the behaviour of the light elements to future work.

\citet{2007arXiv0709.4240D} suggested that the shift in the [C/H] and [N/H] values at first dredge-up requires that the average CEMP has a depth of 0.2\ms\ of accreted material. They point out that this can either be from accreting 0.2\ms\ of material that remains unmixed, or from accreted material mixing to a depth of 0.2\ms. This would imply that, on average, no more than about 0.01\ms\ could be accreted from a 1\ms\ companion (this value rises slightly to around 0.05\ms\ when gravitational settling is taken into account), with this value decreasing rapidly as the companion mass increases. This assumes that thermohaline mixing is as efficient as is presented in this work. If thermohaline mixing is less efficient, then a greater amount of material could be accreted.

\begin{figure}
\includegraphics[width=\columnwidth]{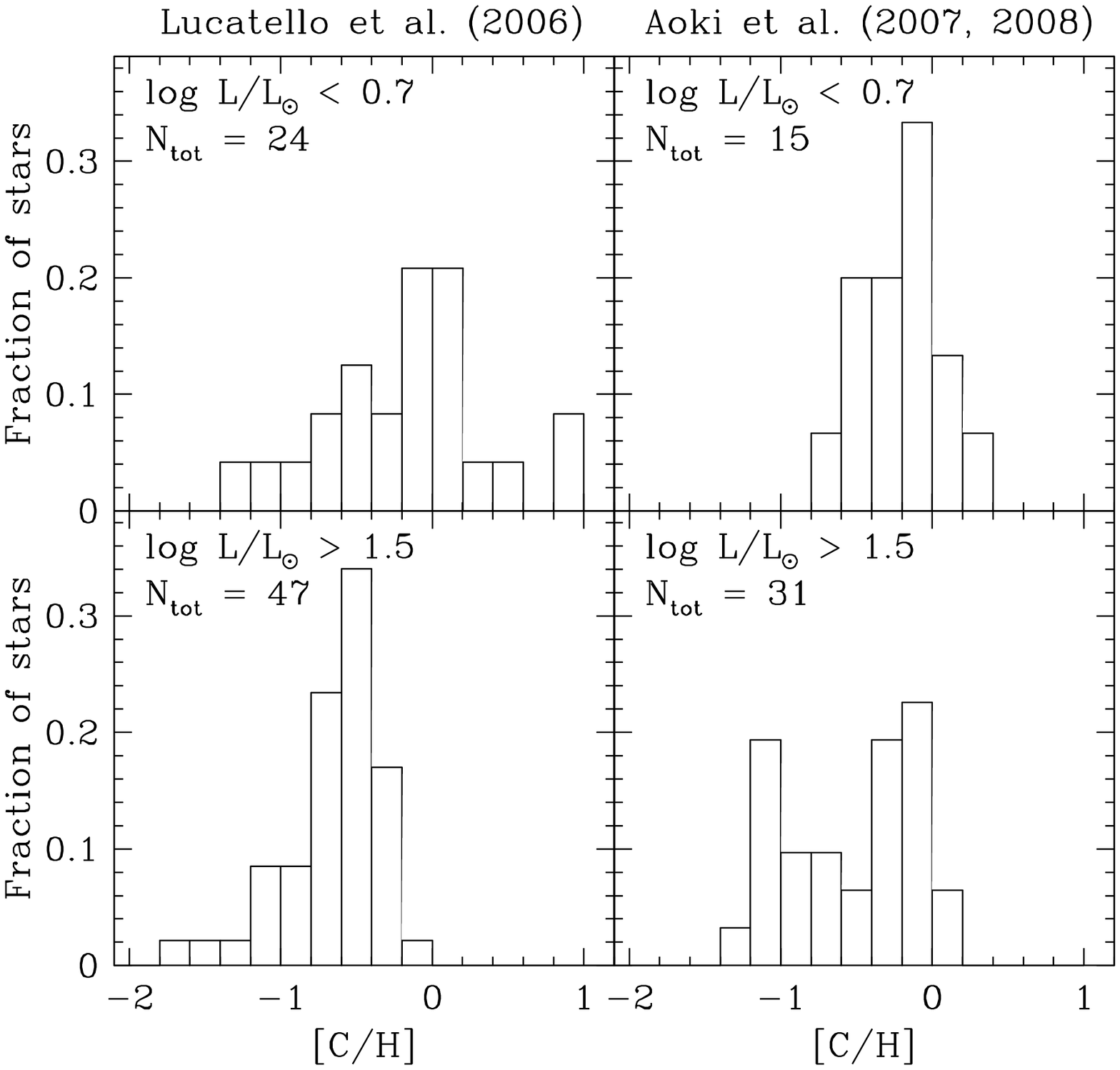}
\caption{The distribution of [C/H] for turn-off stars and giant stars for the data of \citet{2006ApJ...652L..37L} and the combined data of \citet{2007ApJ...655..492A} and \citet{2008ApJ...678.1351A}. Stars are identified as turn-off objects if $\log L/\mathrm{L}_\odot<0.7$ (i.e. they have not yet undergone first dredge-up) and are displayed in the upper panels. Stars are defined as post-first dredge-up giants if $\log L/\mathrm{L}_\odot>1.5$ and are displayed in the lower panels. The total number of stars, $\mathrm{N_{tot}}$, in each plot is also displayed.}
\label{fig:CEMPhistogram}
\end{figure}

\citet{2008ApJ...678.1351A} have suggested that the [C/H] values measured for turn-off stars represent the composition of the donor AGB star. This cannot be reconciled with models including thermohaline mixing, as mixing leads to a large decrease in the [C/H] value. For the star to retain the composition of the AGB donor a substantial amount of material would have to have been accreted from the primary, such that the accreted material dominates the original stellar material. However, the accretion and subsequent thermohaline mixing of a large quantity of material does not agree with the observed differences in distribution of [C/H] from turn-off to giant stars (see Figure~\ref{fig:CEMPhistogram}, right-hand column). Efficient thermoahline mixing of a large quantity of accreted material leads to very little change in [C/H] at first dredge-up. This cannot be the case in the majority of CEMP stars. In addition, if [C/H]$\ \approx 0$ does represent the composition of the donor then there is a major problem with the AGB models which predict substantially larger values. This is not improbable given how little is known about the mass-loss rate at low metallicity and the efficiency of third dredge-up.

The existence of high [C/H] CEMPs suggests that thermohaline mixing is not efficient in all cases. The difference in the distribution of [C/H] for turn-off and giant stars in the sample of \citet{2008ApJ...678.1351A} suggests that some, but not all, turn-off stars with [C/H]$\ \approx0$ do not become mixed after first dredge-up while some do (see Figure~\ref{fig:CEMPhistogram}). After first dredge-up there is still a large proportion of stars with [C/H]$\ \approx 0$ which suggests that little dilution of material has taken place in these objects during first dredge-up.  These objects must have experienced efficient thermohaline mixing, while the others remained unmixed to some degree until first dredge-up. There is less evidence for this phenomenon in the data set of \citet{2006ApJ...652L..37L}, which displays a distinct shift in the peak [C/H] value between turn-off and giant stars (see the left-hand column of Figure~\ref{fig:CEMPhistogram}). In order to predict the distributions of [C/H] for turn-off stars and giants based on the models presented herein, one would need to do a population synthesis calculation for each model set. Attempts at modelling the CEMP population are currently under way \nocite{Izzardetal07inprep}(Izzard et al., in prep).

If thermohaline mixing cannot be inhibited by gravitational settling in the case that a large quantity of material is accreted, we must find another way to reduce its efficiency. A possible mechanism to do this has been put forward by \citet{2007arXiv0708.3864D}. These authors point out that rotationally-induced horizontal turbulent diffusion can suppress thermohaline convection and examine the specific case of mixing on the giant branch \citep[see e.g.][]{2006Sci...314.1580E, 2007A&A...467L..15C}. The molecular weight inversion found there is at least two orders of magnitude smaller than in the case of accretion in CEMPs, so one presumes the turbulent diffusion would need to be much stronger to suppress thermohaline mixing in this situation. The interaction between thermohaline mixing and rotation is an intriguing possibility and merits further study. We are unable to investigate this at present as our code does not include rotational physics.

\section{Conclusions}
We have modelled the accretion of AGB material on to low-mass stars in order to determine what chemical signatures may be observed in CEMP stars. We have examined three specific cases: canonical evolution including only convective mixing, the inclusion of thermohaline mixing and the inclusion of both thermohaline mixing and gravitational settling. We find that thermohaline mixing can lead to accreted material being mixed with between 16 and 88\% of the accreting star, depending on the mass and composition of the accreted material. When gravitational settling is included, thermohaline mixing is severely inhibited when only a small amount of material (around a few $10^{-3}$\ms) is accreted because of the presence of a $\mu$-barrier formed by the settling of helium. This barrier is less and less effective as the amount of accreted material is increased. It is also less effective for more massive companions as the barrier has had less time to be established before accretion on to the secondary occurs.

Models without thermohaline mixing produce turn-off objects with [C/H] values that are too high to match observations. Unless a substantial quantity of material has been accreted, these models also suffer from too much dilution at first dredge-up. Models with thermohaline mixing cannot reproduce the highest [C/H] values observed and predict the existence of low [C/H] objects which are not observed. In order to reproduce the highest [C/H] values, which seem to require the accretion of a large quantity of material from a companion, some alternative mechanism to suppress thermohaline convection is required. The inclusion of gravitational settling can solve the problem of the low [C/H] objects at low luminosity as the $\mu$-barrier prevents mixing when the amount of accreted material is small. However, the inclusion of this physics presents another serious problem namely that carbon will settle from the surface during the main sequence, making it extremely difficult to form turn-off stars with [C/H] values of -1--0.

There also appears to be a problem with the abundance predictions of the AGB models. They do not predict substantial enhancements of both carbon {\it and} nitrogen at the same time (except in a very narrow mass range). Further work needs to be done to find a mechanism capable of producing the observed abundance patterns.

\section{Acknowledgements}
The authors thank Sara Lucatello for making her data available and for helpful discussions. They also thank the referee, Zhanwen Han, for useful comments which have improved the manuscript. RJS is now funded by the Australian Research Council's Discovery Projects scheme under grant DP0879472. He is grateful to Churchill College for the Junior Research Fellowship which has supported him for the last three years. EG thanks the NWO for funding.

\bibliography{../../../masterbibliography}

\appendix

\section{Stellar yields}
Here we present the final gross yields (i.e. the total mass of each isotope ejected) from each of the AGB models for all of the isotopes in the nucleosynthesis network.

\begin{table*}
\begin{scriptsize}
\begin{center}
\begin{tabular}{ccccccccccc}
\hline
Initial & \el{1}{H} & \el{4}{He} & \el{12}{C} & \el{14}{N} & \el{16}{O} & \el{20}{Ne} & & & Mass lost & Final age \\
mass (\ms) & & & & & & & & & (\ms) & (yrs) \\
\hline
1 & 2.76$\times10^{-1}$ & 1.06$\times10^{-1}$ & 1.06$\times10^{-3}$ & 1.28$\times10^{-5}$ & 7.29$\times10^{-5}$ & 3.90$\times10^{-6}$ & & & 3.84$\times10^{-1}$  & 5.83$\times10^{9}$ \\
1.5 & 6.02$\times10^{-1}$ & 2.58$\times10^{-1}$ & 1.06$\times10^{-3}$ & 2.43$\times10^{-5}$ & 4.77$\times10^{-4}$ & 9.72$\times10^{-6}$ & & & 8.72$\times10^{-1}$ & 1.63$\times10^{9}$ \\
2 & 8.96$\times10^{-1}$ & 4.04$\times10^{-1}$ & 2.14$\times10^{-2}$ & 3.44$\times10^{-5}$ & 6.32$\times10^{-4}$ & 2.11$\times10^{-5}$ & & & 1.32$\times10^{0}$ & 7.48$\times10^{8}$ \\
2.5 & 1.19$\times10^{0}$ &  4.91$\times10^{-1}$ & 2.11$\times10^{-2}$ & 3.48$\times10^{-5}$ & 6.53$\times10^{-4}$ & 2.72$\times10^{-5}$ & & & 1.71$\times10^{0}$ & 4.22$\times10^{8}$ \\
3 & 1.53$\times10^{0}$ &  6.11$\times10^{-1}$ &  1.24$\times10^{-2}$ &  7.94$\times10^{-4}$ &  5.33$\times10^{-4}$ &  2.61$\times10^{-5}$ & & & 2.15$\times10^{0}$ &  2.70$\times10^{8}$ \\
3.5 & 1.80$\times10^{0}$ &  8.25$\times10^{-1}$ &  5.53$\times10^{-3}$ &  1.74$\times10^{-2}$ &  6.85$\times10^{-4}$ &  3.82$\times10^{-5}$ & & & 2.65$\times10^{0}$ & 1.88$\times10^{8}$ \\
\hline
Initial mass & \el{2}{H} & \el{3}{He} & \el{7}{Li} & \el{7}{Be} & \el{11}{B} & \el{13}{C} & \el{14}{C} & \el{15}{N} & \el{17}{O} & \el{18}{O} \\
mass (\ms) & & & & & & & & & & \\
\hline
1 & 8.49$\times10^{-16}$ & 1.67$\times10^{-4}$ & 1.04$\times10^{-11}$ & 3.23$\times10^{-11}$ & 1.65$\times10^{-12}$ & 8.64$\times10^{-7}$ & 5.61$\times10^{-8}$ &   8.87$\times10^{-9}$ & 1.79$\times10^{-7}$ & 3.07$\times10^{-8}$ \\
1.5 & 1.15$\times10^{-16}$ & 2.35$\times10^{-4}$ & 1.56$\times10^{-11}$ & 3.32$\times10^{-11}$ & 4.59$\times10^{-12}$ & 8.65$\times10^{-7}$ &  3.81$\times10^{-7}$ &  2.31$\times10^{-8}$ & 7.73$\times10^{-7}$ & 8.33$\times10^{-8}$ \\
2 & 2.06$\times10^{-14}$ & 2.22$\times10^{-4}$ & 1.56$\times10^{-11}$ & 5.22$\times10^{-11}$ &  7.03$\times10^{-11}$ & 9.60$\times10^{-7}$ &  8.80$\times10^{-7}$ &  1.11$\times10^{-8}$ & 1.99$\times10^{-6}$ & 1.03$\times10^{-7}$ \\
2.5 & 8.37$\times10^{-15}$ & 2.45$\times10^{-4}$ &  3.73$\times10^{-12}$ & 1.51$\times10^{-8}$ & 7.08$\times10^{-11}$ & 2.70$\times10^{-6}$ &  5.06$\times10^{-7}$ &  1.68$\times10^{-8}$ &  1.62$\times10^{-6}$ & 1.44$\times10^{-7}$ \\
3 & 6.60$\times10^{-11}$ &  9.25$\times10^{-5}$ &  7.39$\times10^{-11}$ &  3.54$\times10^{-7}$ &  1.91$\times10^{-12}$ &  1.18$\times10^{-3}$ &  8.19$\times10^{-8}$ &  2.86$\times10^{-8}$ &  2.63$\times10^{-6}$ &  3.96$\times10^{-8}$ \\
3.5 & 9.52$\times10^{-17}$ &  3.25$\times10^{-6}$ &  1.71$\times10^{-12}$ &  8.49$\times10^{-10}$ &  3.20$\times10^{-11}$ &  3.38$\times10^{-04}$ &   4.85$\times10^{-9}$ &  5.63$\times10^{-7}$ &  4.16$\times10^{-5}$ &  2.59$\times10^{-9}$ \\ 
\hline
Initial mass & \el{19}{F} & \el{21}{Ne} & \el{22}{Ne} & \el{22}{Na} &\el{23}{Na} & \el{24}{Mg} & \el{25}{Mg} & \el{26}{Mg} & \el{26}{Al}$^m$ & \el{26}{Al}$^g$ \\
mass (\ms) & & & & & & & & & & \\
\hline
1 & 7.87$\times10^{-8}$ & 9.86$\times10^{-9}$ & 1.79$\times10^{-5}$ & 2.24$\times10^{-13}$ & 3.03$\times10^{-7}$ & 1.05$\times10^{-6}$ & 2.05$\times10^{-7}$ &   1.89$\times10^{-7}$ &  0.00 &  3.08$\times10^{-10}$ \\
1.5 & 7.70$\times10^{-7}$ & 8.26$\times10^{-8}$ &  3.33$\times10^{-4}$ & 5.01$\times10^{-13}$ & 4.44$\times10^{-6}$ & 3.32$\times10^{-6}$ & 3.49$\times10^{-6}$ &  1.95$\times10^{-6}$ & 0.00 & 1.03$\times10^{-9}$ \\
2 & 1.05$\times10^{-6}$ & 7.13$\times10^{-7}$ & 1.09$\times10^{-3}$ & 4.78$\times10^{-12}$ & 2.19$\times10^{-5}$ & 1.13$\times10^{-5}$ & 3.24$\times10^{-5}$ &   2.87$\times10^{-5}$ &  0.00 &  2.81$\times10^{-9}$ \\
2.5 & 3.18$\times10^{-7}$ &  9.63$\times10^{-7}$ &  6.97$\times10^{-4}$ &  7.71$\times10^{-12}$ &  1.57$\times10^{-5}$ & 1.26$\times10^{-5}$ &  5.21$\times10^{-5}$ &  9.36$\times10^{-5}$ &  0.00 &  2.47$\times10^{-9}$ \\
3 & 7.31$\times10^{-8}$ &  3.60$\times10^{-7}$ &  1.94$\times10^{-4}$ &  9.34$\times10^{-10}$ &  6.28$\times10^{-6}$ &  8.21$\times10^{-6}$ &  2.75$\times10^{-5}$ &   7.07$\times10^{-5}$ &  0.00 &  2.39$\times10^{-9}$ \\
3.5 & 5.68$\times10^{-9}$ &  3.70$\times10^{-7}$ &  1.57$\times10^{-4}$ &  3.59$\times10^{-7}$ &  1.56$\times10^{-4}$ &  1.39$\times10^{-5}$ &  7.20$\times10^{-5}$ &  2.00$\times10^{-4}$ &  0.00 &  3.28$\times10^{-7}$ \\
\hline
Initial mass & \el{27}{Al} & \el{28}{Si} & \el{29}{Si} & \el{30}{Si} & \el{31}{P} & \el{32}{S} & \el{33}{S} & \el{34}{S} & \el{56}{Fe} & \el{57}{Fe} \\
mass (\ms) & & & & & & & & & & \\
\hline
1 & 1.12$\times10^{-7}$ & 1.26$\times10^{-6}$ & 6.64$\times10^{-8}$ & 4.57$\times10^{-8}$ & 1.62$\times10^{-7}$ & 7.64$\times10^{-7}$ & 6.29$\times10^{-9}$ &  3.59$\times10^{-8}$ &  2.24$\times10^{-6}$ & 5.56$\times10^{-8}$ \\
1.5 & 2.61$\times10^{-7}$ & 2.85$\times10^{-6}$ & 1.54$\times10^{-7}$ & 1.08$\times10^{-7}$ & 3.93$\times10^{-7}$ & 1.72$\times10^{-6}$ & 1.49$\times10^{-8}$ &   8.01$\times10^{-8}$ & 5.00$\times10^{-6}$ & 1.34$\times10^{-7}$ \\
2 & 6.07$\times10^{-7}$ & 4.42$\times10^{-6}$ & 2.59$\times10^{-7}$ & 1.85$\times10^{-7}$ & 6.32$\times10^{-7}$ & 2.59$\times10^{-6}$ & 2.30$\times10^{-8}$ &  1.19$\times10^{-7}$ &  7.40$\times10^{-6}$ & 2.04$\times10^{-7}$  \\ 
2.5 & 1.65$\times10^{-6}$ &  6.41$\times10^{-6}$ &  4.64$\times10^{-7}$ &  3.47$\times10^{-7}$ &  1.01$\times10^{-6}$ &  3.38$\times10^{-6}$ &  2.98$\times10^{-8}$ &   1.55$\times10^{-7}$ &  9.62$\times10^{-6}$ &  2.47$\times10^{-7}$ \\
3 & 1.74$\times10^{-6}$ &  7.86$\times10^{-6}$ &  5.60$\times10^{-7}$ &  4.58$\times10^{-7}$ &  1.29$\times10^{-6}$ &  4.34$\times10^{-6}$ &  3.75$\times10^{-8}$ &  2.00$\times10^{-7}$ &  1.24$\times10^{-5}$ &  3.09$\times10^{-7}$ \\
3.5 & 4.18$\times10^{-6}$ &  1.12$\times10^{-5}$ &  1.04$\times10^{-6}$ &  9.34$\times10^{-7}$ &  2.61$\times10^{-6}$ &  5.61$\times10^{-6}$ &  5.32$\times10^{-8}$ &  2.47$\times10^{-7}$ &  1.51$\times10^{-5}$ &  3.75$\times10^{-7}$ \\
\hline
Initial mass & \el{58}{Fe} & \el{59}{Fe} & \el{60}{Fe} & \el{59}{Co} & \el{58}{Ni} &
\el{59}{Ni} & \el{60}{Ni} & \el{61}{Ni} & g & \\
mass (\ms) & & & & & & & & & & \\
\hline
1 & 8.86$\times10^{-9}$ & 6.59$\times10^{-16}$ & 2.53$\times10^{-12}$ & 6.85$\times10^{-9}$ & 9.49$\times10^{-8}$ & 3.55$\times10^{-11}$ & 3.76$\times10^{-8}$ &  1.71$\times10^{-9}$ & 5.36$\times10^{-10}$  \\
1.5 & 4.64$\times10^{-8}$ & 1.36$\times10^{-14}$ & 1.72$\times10^{-10}$ & 2.13$\times10^{-8}$ & 2.09$\times10^{-7}$ & 3.16$\times10^{-10}$ &  8.35$\times10^{-8}$ &   4.52$\times10^{-9}$ & 6.22$\times10^{-9}$ \\
2 &  1.25$\times10^{-7}$ & 3.16$\times10^{-11}$ &  1.11$\times10^{-8}$ &  5.18$\times10^{-8}$ & 3.08$\times10^{-7}$ & 2.09$\times10^{-10}$ & 1.22$\times10^{-7}$ &  6.24$\times10^{-9}$ & 2.25$\times10^{-8}$ \\
2.5 & 8.85$\times10^{-8}$ &  4.90$\times10^{-10}$ &  5.23$\times10^{-8}$ &  4.85$\times10^{-8}$ &  4.05$\times10^{-7}$ &  1.52$\times10^{-10}$ &  1.60$\times10^{-7}$ &  7.43$\times10^{-9}$ & 4.47$\times10^{-8}$ \\ 
3 & 6.92$\times10^{-8}$ &  5.83$\times10^{-10}$ &  3.30$\times10^{-8}$ &  4.70$\times10^{-8}$ &  5.24$\times10^{-7}$ &  1.08$\times10^{-10}$ &  2.07$\times10^{-7}$ &  9.39$\times10^{-9}$ & 2.86$\times10^{-8}$ \\
3.5 & 7.15$\times10^{-8}$ &  1.87$\times10^{-10}$ &  5.35$\times10^{-8}$ &  5.12$\times10^{-8}$ &  6.40$\times10^{-7}$ &  9.91$\times10^{-11}$ &   2.53$\times10^{-7}$ &  1.13$\times10^{-8}$ & 1.27$\times10^{-7}$ \\ 
\hline
\end{tabular}
\end{center}
\end{scriptsize}
\caption{Final gross yields in stellar masses for all the isotopes in the nucleosynthesis network.}
\label{tab:agbyields}
\end{table*}

\label{lastpage}

\end{document}